\documentstyle[prl,aps,preprint,eqsecnum,epsfig]{revtex}

\begin{document}
 
\title{Impact of a Multi-TeraFlop Machine to Gravitational Physics}
 
\author{W.\ M.\ Suen}
 
\address{McDonnell Center for the Space Sciences, Department of 
Physics, Washington University, \\
St Louis, MO 63130, U S A}
 
\date{\today}
 
\maketitle

\begin{abstract}
 
A multi-TeraFlop/TeraByte machine will enable the application of the
Einstein theory of gravity to {\it realistic}
astrophysical processes.  Without the computational power, the
complexity of the Einstein theory restricts most studies based on it
to the quasi static/linear near-Newtonian regime of the theory.

The application of the Einstein theory to realistic astrophysical
processes is bound to bring deep and far-reaching scientific
discoveries, and produce results that will inspire the general
public.  It is an essential component in developing the new frontier of
gravitational wave astronomy -- an exciting new window to observe our
universe.

The computational requirements of carrying out numerical simulations based
on the Einstein theory is discussed with an explicit example,
the coalescence of a neutron star binary.

This document is prepared for presentation at the National
Computational Science Alliance User Advisory Council Meeting at NSF,
June 1998, in support of the funding of a NSF TeraFlop computer.

\end{abstract}
 
\draft
 
\pagebreak

\section{Introduction - Astronomy of the Next Century and
Gravitational Physics}

Two major directions of astronomy in the next century are {\em high
energy ($x$-ray, $\gamma$-ray) astronomy} and {\em gravitational wave
astronomy.} The former is driven by observations by $x$- and
$\gamma$-ray satellites, e.g., CGRO, AXAF, XTE, HETE II,
GLAST~\cite{missions}, current or planned for the next few years.
High energy radiation is often emitted in regions of strong
gravitational fields, near black holes (BHs) or neutron stars (NSs).
One of the biggest mysteries of modern astronomy, $\gamma$-ray bursts,
is likely to be generated by events involving NSs or BHs.  For the full
description of strong, dynamic gravitational fields, we need
Einstein's theory of general relativity.

The second major direction, gravitational wave astronomy, involves
directly the dynamical nature of spacetime in the Einstein theory of
gravity.  The tremendous recent interest in this frontier is driven by
the gravitational wave observatories presently being built or planned
in US, Europe and outer space, e.g., LIGO, VIRGO, GEO600, LISA, LAGOS
~\cite{nasa}, and the Lunar Outpost Astrophysics Program~\cite{nasa}.
The American LIGO and its European counterparts VIRGO and GEO600 are
scheduled to be on line in a few years~\cite{LIGOweb}, making
gravitational wave astronomy a reality.  These observatories provide a
completely new window on the universe: existing observations are
mainly provided by the electromagnetic spectrum, emitted by individual
electrons, atoms or molecules, easily absorbed, scattered and
dispersed.  Gravitational waves are produced by coherent bulk motion
of matter and travel nearly unscathed through space, coming to us
carrying the information of the strong field regions where they were
originally generated.~\cite{thorne96} This new window will provide very
different information about our universe that is either difficult or
impossible to obtain by traditional means.

The numerical determination of the gravitational waveform is crucial
for gravitational wave astronomy.  Physical information in the data is
to be extracted through template matching techniques~\cite{Cutler93},
which {\em presupposes} that reliable example waveforms are
known.~\cite{template} Gravitational waveforms are important both as
probes of the fundamental nature of gravity, and for the unique
physical and astronomical information they carry.  The information
would be difficult to obtain otherwise, ranging from nuclear physics
(e.g., the EOS of NSs~\cite{Cutler93}) to cosmology (e.g., direct
determination of the Hubble constant~\cite{schutz86}).  In most situations, the
gravitational waveforms cannot be calculated without full scale
general relativistic numerical simulations.

In short, both of these frontiers of astronomy call for
numerical simulations based on the Einstein theory of gravity.
If astrophysicists are to fully understand the non-linear and
dynamical gravitational fields involved in these observational data,
detailed modeling taking dynamic general relativity into full account
must be carried out.  

\section{Challenges of Computational General Relativistic Astrophysics}

The application of the Einstein theory of gravity to
{\it realistic} astrophysical systems needs computational power in the
range of (at least) multi-TeraFlop/TeraByte, and corresponding
capabilities in visualization, networking and storage.

\noindent $\bullet$ Computational challenges due to the complexity of
the physics involved: The Einstein equations are probably the most
complex partial differential equations in all of physics, forming a
system of dozens of coupled, nonlinear equations, with thousands of
terms, of mixed hyperbolic, elliptic, and even undefined types in a
general coordinate system.  The evolution has elliptic constraints
that should be satisfied at all times.  In simulations without
symmetry, as would be the case for realistic processes, it involves
hundreds of 3D arrays, and ten of thousands of operations per grid
point per update.  Moreover, for simulations of
astrophysical processes, we need to integrate numerical relativity
with traditional tools of computational astrophysics, including
hydrodynamics, nuclear astrophysics, radiation transport and
magneto-hydrodynamics, which govern the evolution of the source terms
(i.e., the right hand side) of the Einstein equations.  This
complexity demands massively parallel computation.

\noindent $\bullet$ The object under numerical construction being the
spacetime itself presents unique challenges: According to the
singularity theorems of general relativity, region of strong gravity
often generate spacetime singularities.  Due to the need to avoid
spacetime singularities~\cite{numrel}, and to obtain long term
stability in the numerical simulations, sophisticated control of the
coordinate system is needed for the construction of a
numerical spacetime.  This dynamic interplay between the
spacetime being constructed and the computational coordinate choice
itself (``gauge choice'') is a unique feature of general relativity
that makes the numerical simulations much more demanding.  Beside
extra computational power, advanced visualization tools, preferably
real time interactive ``window into the oven'' visualization, are
particularly useful in the numerical construction.

\noindent $\bullet$ The multi-scale problem: Astrophysics of strongly
gravitating systems inherently involves many length and time scales.
The microphysics of the shortest scale (the nuclear force), controls
macroscopic dynamics on the stellar scale, such as the formation and
collapse of neutron stars (NSs).  On the other hand, the stellar scale
is at least 10 times {\it less} than the wavelength of the
gravitational waves emitted, and many orders of magnitude less than
the astronomical scales of their accretion disk and jets; these larger
scales provide the directly observed signals.  Numerical studies of
these systems, aiming at direct comparison with observations,
fundamentally require the capability of handling a wide range of
dynamical time and length scales.  While such multi-scale problems can
be handled with advanced 3D AMR techniques, it leads to further
requirements on computation power and (3D AMR) visualization.

In short, in order to meet the challenges of Computational General
Relativistic Astrophysics we need to push not only the frontier of the
computation power for number crunching.  The visualization requires
basically as much computer power as what generates the data.  The
highly multi-disciplinary nature of the research demands collaborative
code development.  The large amount of data, visualization needs, and
collaborative effort require high performance networking and
meta-computing.  In the following section we use a specific sample
problem to illustrate the requirements on Flop rate, memory, disk and
storage sizes, which in turns determine the base line of visualization
and networking requirements.  Where we stand at present will also be
discussed briefly.

\section{Neutron Star Coalescence As An Example on Computational Requirements}

We use the problem of coalescing binary neutron stars to show the
computational requirements in general relativistic astrophysics.  The
reason that the coalescence of neutron stars is a meaningful example
is many-fold: It is a significant problem in astrophysics and
astronomy; it involves many ingredients in general relativistic
astrophysics;  and it is a problem attracting much current research
effort both nationally and internationally.

\medskip

\noindent $\bullet$ Coalescing neutron star binary systems are common in the
Universe, with the well known Hulse-Taylor binary pulsar PSR1913+16
being an example.  The coalescence events are expected
to be detectable by LIGO, with an observation rate of 29 yr$^{-1}$ for
$h=0.5$ and 43 yr$^{-1}$ for $h=0.8$. ~\cite{finn95}

\noindent $\bullet$ The physical information
in LIGO data is to be extracted through the standard template matching
technique~\cite{Cutler93}.  For this we need to determine the waveforms 
of the gravitational radiation generated by the coalescence events,
which can only be obtained through large scale simulations.

\noindent $\bullet$ A very enticing reason for studying the
coalescence event lies in the fact that observations of such events by
gravitational wave observatories may allow us to determine
cosmological parameters like $H_0$ and $q_0$, without going through the
cosmic distance ladder, and is independent of the optical
identification of the source and the evolution of the source rate
density with redshift.\cite{schutz86,finn95}

\noindent $\bullet$ Gravitational wave signals from coalescing
binaries may reveal important information on the equation of state of
dense nuclear matter, including the nuclear compression modulus, the
hadronic effective masses, the relative hyperon-nucleon and
nucleon-nucleon coupling constants, possible kaon condensation and a
quark/hadron phase transition.

\noindent $\bullet$  Study of coalescing neutron star 
binaries may also answer other long standing questions in nuclear
astrophysics. 
NSNS binary mergers may eject
\cite{ls76} extremely neutron-rich matter which decompresses,
beta-decays and neutron captures, forming the classical r-process
\cite{lmrs77,mbc92}.  Detailed numerical simulations of
the shock heating and mass ejection process are needed.

\noindent $\bullet$ Coalescing neutron star binaries are among the
most popular candidates of gamma ray bursts.~\cite{paczyn86,piran95}
In order to evaluate the feasibility of the model, detailed studies
taking the full general relativistic effects are needed to determine the
maximum possible energy released, heating and mass ejection in the
coalescence process.

\subsection{Minimum Configuration}

\noindent $\bullet$ Description of the Physical System:

Two 1.4 solar mass neutron stars in head-on collision falling in from
infinity.  General relativistic simulation begins when the two stars are
$4 R$ apart, with $R=$ radius of star.  Simulation covers $10ms$ in time for 
the dynamics of the merging and ringdown phases, and $20R$ in space for resolution
and boundary considerations.

\noindent $\bullet$ Purposes:  Study the general relativistic dynamics of the merging and ringdown phases
of head-on collision.

\noindent $\bullet$ Grid Setup: Resolution=25 gridpoints/$R$,  Total Grid Size = $500 ^ 3 = 10 ^ 8 $

\noindent $\bullet$ Memory Requirement: $180GBytes$

\noindent $\bullet$ Floating Point Operations:

Flops/gridpt/time step = $10 ^4 ~~~~$ (With only weak coordinate control)

Total number of time steps = $10 ^4 ~~~~$

Total flops =  $10 ^ {16} ~~~~$

Run time = 3 hours   ~~~~ (With 1 TeraFlops sustained)

\noindent $\bullet$ Disk:

Run time disk size = 800 GBytes ~~(Output 10 functions with 1/100 sampling)

Storage = 8 TeraBytes ~~~~(with 10 runs for comparison studies)

\noindent $\bullet$ Present Status:

Code for carrying out this simulation is {\it currently} available.  A
code constructed for the NASA Neutron Star Grand Challenge Project which is
capable of solving the full Einstein equations coupled to general
relativistic hydrodynamics has recently been released.~\cite{ourwebpage} This
code has been tested on a 1024 node T3E-1200 (provided for
the neutron star project for performance tests, though not available for
production runs), achieving 142GFLops and linear scaling up to 1024 nodes.
A summary of the test results are given below.
(The NSF Black Hole Grand Challenge Project is also constructing
massively parallel code for solving the Einstein equations, see
~\cite{matzner} for present status.)

\noindent {\em \underline{Code tested}}: 
NASA Neutron Star Grand Challenge GR3D Einstein Spacetime (ADM)
coupled to MAHC HYPERBOLIC\_HYDRO  (code tested with the released
version, without special tuning for this 1024 node machine.)

\noindent {\em \underline{Date tested}}: May 10, 1998 

\begin{verbatim}
                                  32 bit             64 bit 
--------------------------------------------------------------------
Grid Size per Processor           84x84x84           66x66x66 
Processor topology                8 x8 x16           8 x8 x16 
Total Grid Size                   644 x 644 x 1284   500 x 500 x 996 
Single Proc MFlop/sec             144.35             118.33 
Aggregate GFlop/sec               142.2              115.8 
Scaling efficiency                96.2%              95.6% 
--------------------------------------------------------------------
\end{verbatim}

\subsection{Medium Configuration}

\noindent $\bullet$ Description of Physical System:

Two 1.4 solar mass neturon stars in inspiral coalescence.  Full
general relativistic simulation begins when stars enter the last 8
orbits.  Simulation covers $60ms$ in time and $22R$ in space.

\noindent $\bullet$ Purposes:

Study the general relativistic inspiral dynamics
beginning with the 3PN breakpoint.  This enables reliable initial
data to be set.  Study the effects of the angular momentum and
gravitational radiation backreaction on shock heating in the merger phase.

\noindent $\bullet$ Grid Setup:
Resolution=50 gridpoints/$R$,  Total Grid Size = $10 ^ 9 $

\noindent $\bullet$ Memory Requirement: $1.8 TBytes$

\noindent $\bullet$ Floating Point Operations:

Flops/gridpt/time step = $10 ^4 ~~~~$ (With only weak coordinate control)

Total number of time steps = $10 ^5 ~~~~$

Total flops =  $10 ^ {18} ~~~~$

Run time = 300 hours ~~~~(With 1 TeraFlops sustained)

\noindent $\bullet$ Disk:

Run time disk size = 20 TBytes ~~~(Output 10 functions with 1/400 sampling)

Storage = 100 TeraBytes (with 5 runs for comparison studies)

\noindent $\bullet$ Visualization: Need parallel visualization engine.

\noindent $\bullet$ Present Status:

Code basically ready for pilot studies.  Tests of the effects of the
implementation of weak coordinate control to be performed.

\subsection{Preferred Configuration}

\noindent $\bullet$ Description of Physical System:

Two 1.4 solar mass neturon stars in inspiral coalescence.  Full
general relativistic simulation begins when stars enter the last 8
orbits.  Simulation covers $60ms$ in time and $40R$ in space (one
wavelength for gravitational wave with period 1ms).

\noindent $\bullet$ Purposes:  Study the same system with strong coordinate control and more reliable wavefrom
extraction.

\noindent $\bullet$ Grid Setup:

Resolution=50 gridpoints/$R$,  Total Grid Size = $10 ^ {10} $

\noindent $\bullet$ Memory Requirement: $18 TBytes$

\noindent $\bullet$ Floating Point Operations:

Flops/gridpt/time step = $10 ^5 ~~~~$ (With strong coordinate control)

Total number of time steps = $10 ^5 ~~~~$

Total flops =  $10 ^ {20} ~~~~$

Run time = 3,000 hours ~~~~(With 10 TeraFlops sustained)

\noindent $\bullet$ Disk:

Run time disk size = 200 TBytes ~~(Output 10 functions with 1/400 sampling)

Storage = 1000 TeraBytes 

(with 5 runs for comparison studies, template preparation not included)

\noindent $\bullet$ Need to push the frontiers on computation, storage, visualization, and networking.

\noindent $\bullet$ Present Status:

Code basically ready for pilot studies.  Efficient control
the coordinate system to be investigated.

\section{Acknowledgements}

I thank S. Finn, K. Blackburn, M. Miller, L. Smarr, B. Sugar,
M. Tobias, J. Towns, C. Will, and J. York for useful input in
preparing this document.

The gereral relativistic astrophysics code "GR3D" discussed in Sec. 4
is developed by the NCSA-Potsdam-Wash U numerical relativity
collaboration, with support from the NSF Gravitational Physics Program
Grant No. Phy-96-00507, NASA HPCC/ESS Grand Challenge Applications
Grant No. NCCS5-153, NSF NRAC Allocation Grant no. MCA93S025, and
the Albert Einstein Institute.




\end{document}